\documentstyle[11pt]{article}
\def\pri{^{\, \prime}}

\def\np#1{{\em Nucl.~Phys.}~{\bf B#1}\ }
\def\deg{\ifmmode{^{\circ}}\else ${^{\circ}}$\fi}

\def\itm#1{\item[$(#1)$]}
\def\gsim{\,\raisebox{-0.13cm}{$\stackrel{\textstyle>}{\textstyle\sim}$}\,}
\def\lsim{\,\raisebox{-0.13cm}{$\stackrel{\textstyle<}{\textstyle\sim}$}\,}
\def\bi{\begin{itemize}}
\def\ei{\end{itemize}}
\def\ed{\end{document}}
\def\be{\begin{equation}}
\def\ee{\end{equation}}
\def\bea{\begin{eqnarray}}
\def\eea{\end{eqnarray}}
\def\req#1{(\ref{eq:#1})}
\def\eq#1{Eq.~(\ref{eq:#1})}
\def\labeq#1{\label{eq:#1}}

\def\tfrac#1#2{{\textstyle\frac{#1}{#2}}}

\def\thalf{\tfrac{1}{2}}

\def\gev{\ \mbox{GeV}}

\def\index{n_{\rm ref}}
\def\msqeff{m^2_{\rm eff}}
\def\tdel{{\tilde \Delta}}
\def\tsig{{\tilde \sigma}}
\def\gsim{\raisebox{-0.5ex}{$\stackrel{>}{\sim}$}}
\def\lsim{\raisebox{-0.5ex}{$\stackrel{<}{\sim}$}}

\def\eb{\end{thebibliography}}

\def\nn{\nonumber}
\def\nns{\nn\\[.1in]}

\def\labeq#1{\label{eq:#1}}
\def\req#1{(\ref{eq:#1})}
\def\eq#1{Eq.~(\ref{eq:#1})}
\def\tr{\ifmmode{\mbox{Tr}}\else Tr\fi}

\def\np{\nu\pri}
\def\mnu{m_{\nu}}
\def\mnus{\mnu^2}
\def\nuth{\nu_{th}}
\def\ov#1{\overline{#1}}
\def\cs{cross section}
\def\si{strong interaction}
\def\cl{charged lepton}
\def\apm{A_{\pm}}

\def\lm{\lambda}

\def\egzk{\ifmmode{E_{GZK}}\else $E_{GZK}$\fi}
\def\es{\ifmmode{E^*}\else $E^*$\fi}
\def\sigs{\ifmmode{\sigma^*}\else $\sigma^*$\fi}
\begin{document}
\begin{titlepage}
\begin{flushright}  {\sl NUB-3191/98-Th}\\
                    {\sl VAND-TH-98-18}\\
{\sl October 1998}\\
hep-ph/9810533
\end{flushright}
\vskip 0.5in

\begin{center}
{\Large\bf Lower Energy Consequences of an Anomalous\\
           High--Energy Neutrino Cross--Section}\\
[.5in]
{Haim Goldberg$^1$ and T. J. Weiler$^2$}\\[.1in]
$^1${\it Department of Physics, Northeastern University, Boston, MA 02115,
USA}\\
$^2${\it Department of Physics \& Astronomy, Vanderbilt University,
Nashville, TN 37235, USA}
\end{center}
\vskip 0.4in

\begin{abstract}
A new strong--interaction has been postulated for neutrinos above $\sim
10^{19}$ eV to explain the production of highest--energy cosmic ray events. We
derive a dispersion relation relating the hypothesized
high--energy cross--section to
the lower--energy neutrino--nucleon elastic amplitude. Remarkably, we find
that the real forward amplitude becomes anomalous seven orders of magnitude
lower in energy than does the total cross--section. We discuss possible
measurable consequences of this early onset of new neutrino physics, and
conclude that a significantly enhanced elastic $\nu N$ scattering rate may
occur for the neutrino beams available at Fermilab and CERN.
\end{abstract}
\end{titlepage}
\setcounter{page}{2}

\section{Introduction}
The discoveries by the AGASA \cite{akeno}, Fly's Eye \cite{eye}, Haverah Park
\cite{hp}, and Yakutsk \cite{yak} collaborations of air shower events with
energies above the Greisen--Zatsepin--Kuzmin (GZK) cutoff of $\sim 5\times
10^{19}$ eV challenge the Standard Model (SM) of particle physics and the hot
big--bang model of cosmology. Not only is the mechanism for particle
acceleration to such extremely high energy cosmic rays (EHECRs) controversial,
but also the propagation of EHECRs over cosmic distances is problematic. Above
the resonant threshold for $\Delta^*$ production, $\sim 5\times 10^{19}$ eV,
protons lose energy by the scattering process
$p+\gamma_{2.7K}\rightarrow\Delta^*\rightarrow N+\pi$; $\gamma_{2.7K}$ denotes
a photon in the $2.7K$ cosmic background radiation. This is the mechanism for
the much--heralded GZK cutoff \cite{gzk}. For every mean free path $\sim 6$ Mpc
of travel, the proton loses 20\% of its energy on average \cite{bzinsky}.   A
proton produced at its cosmic source a distance $D$ away will on
average arrive at earth with only a fraction $\sim (0.8)^{D/6\,{\rm Mpc}}$ of
its original energy.  Of course, proton energy is not lost significantly if the
highest energy protons come from a rather nearby source,
$\lsim$ 50 to 100 Mpc.\footnote
{The suggestion has been made that hot spots of radio
galaxies in the supergalactic plane at distances of tens of megaparsecs may be
the sources of the super--GZK primaries \cite{stanev}. Present statistics are
too limited to validate or invalidate this proposal. }
However, acceleration of protons to $\sim 10^{20}$ eV, if possible at all, is
generally believed to require the most extremely--energetic compact sources,
such as active galactic nuclei (AGNs) \cite{AGNmax} or gamma--ray bursts (GRBs)
\cite{GRBmax}.
Since AGNs and GRBs are
hundreds of megaparsecs away, the energy requirement at an AGN or GRB for a
proton which arrives at earth with a super--GZK energy is unrealistically high
\cite{50Mpc}.
A primary nucleus mitigates the cutoff problem (energy per nucleon is reduced
by 1/A), but above $\sim 10^{19}$ eV nuclei should be photo--dissociated by the
2.7K background \cite{nucleus}, and possibly disintegrated by the particle
density ambient at the astrophysical source. Gamma--rays and neutrinos are
other possible primary candidates for the highest energy events.
The mean free path, however, for a $\sim 10^{20}$ eV photon to annihilate on
the radio background to
$e^+ e^-$ is believed to be only $10$ to $40$ Mpc \cite{bzinsky}.

Concerning the neutrino hypothesis, the Fly's Eye event occurred high in the
atmosphere, whereas the event rate expected in the SM for early development of
a neutrino--induced air shower is down from that of an electromagnetic or
hadronic interaction by six orders of magnitude \cite{hs}. On the other hand,
there is evidence that the arrival directions of some of the
highest--energy primaries are paired with the directions of events lower in
energy by an order of magnitude, and displaced in time by about a year
\cite{AGASApairs}. As noted in \cite{AGASApairs}, such event--pairing argues for
stable neutral primaries coming from a source of considerable duration.
Neutrino primaries do satisfy this criterion.
Furthermore, a recent analysis of the arrival directions of the super--GZK
events offers a tentative claim of a correlation with the directions of
radio--loud quasars \cite{BF98}.  If this correlation is validated with
future data,
then the propagating cosmic particle must be charge neutral and have a
negligible magnetic moment.  The neutrino again emerges as the only candidate
among the known particles.
Thus, it is of some interest to
examine the possibility that the primary cosmic rays above the GZK cutoff
energy are neutrinos, but with some large non--SM cross section that allows
them to initiate air showers high in the atmosphere \cite{SInu}.
To mimic
hadronically--induced air showers, the new neutrino cross section must be of
hadronic strength, $\sim 100$~mb, above \egzk$\equiv 5\times 10^{19}\ \rm{eV}$.

The use of an anomalously large high--energy neutrino--interaction
to model the observed super--GZK events
has been criticized on the ground that the onset of new neutrino physics cannot
be so rapid as to hide the new
interactions from experimental view at lower energies \cite{bgh98}.
In turn, the
criticism of the model has itself been criticized \cite{meanmo},
on the ground that a negative conclusion was drawn from simple perturbative
calculations of single scalar or vector exchange models.
In the present paper we
analyze the hypothesized rapid--rise in the neutrino cross section using
an approach which is completely model-independent.
We assume the hypothesized new physics, whatever
it may be, holds at and above an energy scale
\es, and use dispersion relations to provide a rigorous nonperturbative
calculation of the growth of the {\sl elastic} neutrino amplitude at much
lower energies.
We find that if the
new physics dominates the neutrino {total} \cs\ with a hadronic value $\sigma^*$
above the lab energy \es, then
the real part of the new strong--interaction
elastic amplitude at lower energy $E$ is given by
$\frac{1}{2\pi}\frac{E}{E^*}\sigma^*$.
Thus, the new physics has the possibility of inducing a significant anomalous
contribution to low energy neutrino propagation in matter, and to
low energy {\sl elastic} neutrino--nucleon scattering.

The plan of the paper is as follows:
in the next Section  we examine the implication for the low
energy elastic $\nu N$ amplitude
of an anomalous $\nu N$ cross section at very high energies.
In two subsections we discuss the experimental possibilities for testing this
hypothesis. We calculate the anomalous index of refraction and effective
potential induced for lower energy neutrinos in matter, and find that they are
probably not measurable. We compute the enhanced elastic $\nu N$ scattering
cross--section at lower energies, and find that it may be measurable with
neutrino beams existing at present accelerators. Section 3 contains some
discussion and questions for further study, and our conclusions. The dispersion
relations for the relevant $\nu N$ amplitudes, central to this paper, are
derived in an Appendix.

\section{Low--Energy Elastic Amplitude from High--Energy Threshold}

Suppose that there is a new neutrino-nucleon interaction of hadronic-strength
at neutrino lab energy $E\pri\ge \es \;(i.e. \,\sqrt{s\pri}\ge \sqrt{2M\es}),$
hypothesized to explain the air showers observed above the GZK cutoff.
Then, for $E\pri\ge \es$ we have
\be
\sigma^{\nu N}_{tot}(E\pri,\pm)=\sigs\ \ .
\labeq{sigstar}
\ee
To address the question ``Are there anomalous neutrino interactions below
$\es$?''
we invoke the dispersion relation for neutrino-nucleon scattering
derived in the Appendix:
\be
{\rm Re}\ \apm(E)-{\rm Re}\ \apm(0)=\frac{E}{4\pi}{\cal P}\int_0^{\infty}\
dE\pri\
\left(
\frac{\sigma^{\nu N}_{tot}(E\pri,\pm)}{E\pri(E\pri-E)}\ +\
\frac{\sigma^{\bar \nu N}_{tot}(E\pri,\pm)}{E\pri(E\pri+E)}
\right)\ \ ,
\labeq{dispsubt}
\ee
where $\apm(E)$ are invariant $\nu$-$N$ amplitudes, labeled by the
nucleon helicity and
defined in Eqs.~\req{s} and \req{t} of the Appendix, and $\cal P$ denotes the
principle value of the integral. The new physics dominates the neutrino-nucleon
dispersion integral
\req{dispsubt} for $E\pri\ge E^*.$ Motivated by simplicity and the behavior of
the SM strong--interaction, let us assume that $\sigs$ is independent of
helicity and energy, and that the new hadronic component of the neutrino cross
section obeys the Pomeranchuk theorem:
\be
\sigma^{\nu N}_{tot}(E,\pm)-\sigma^{\bar \nu N}_{tot}(E,\pm)
\stackrel{\nu\rightarrow\infty}{\longrightarrow} 0\ \ .
\labeq{pom}
\ee
These assumptions and the dispersion relation lead directly to an evaluation
of the difference ${\rm Re}\ \apm(E)-{\rm Re}\ \apm(0)$,
which is not calculable in perturbation theory.
Ignoring ordinary weak interaction contributions to the dispersion integral
(which amounts to the omission of electroweak radiative corrections),  we find
for the real part of the amplitude at energy
$E$

\bea
{\rm Re}\ \apm(E)&\simeq&{\rm Re}\ \apm(0) + \Delta(E)\nns
\Delta(E)&\equiv& \frac{1}{2\pi}\frac{E}{E^*}\sigma^*\ \ .
\labeq{delta}
\eea
To arrive at simpler expressions below, and to take advantage of the assumed
helicity--independence of the new interaction, it is
convenient to work with the vector and axial vector amplitudes
\bea
C_V&=&\thalf(A_-+A_+)\nns
C_A&=&\thalf(A_--A_+)\ \ .
\labeq{va}
\eea
{}From \req{delta}, we have (again ignoring weak contributions to the
dispersion integral)
\bea
{\rm Re}\ C_V(E)&=&{\rm Re}\ C_V(0)\ + \
\Delta(E)\nns
{\rm Re}\ C_A(E)&=&{\rm Re}\ C_A(0)\ \ .
\labeq{vadelta}
\eea

The elastic amplitudes at $E=0$ are determined from $Z$-exchange, and are given
by
\bea
C_V(0)&=&-\frac{G_F}{\sqrt{2}}\ [T_3(N)-2\sin^2{\theta_W}Q(N)] \nns
C_A(0)&=&-\frac{G_F}{\sqrt{2}}\ T_3(N)\ \ ,
\labeq{vazero}
\eea
with $N=p$ or $n$.
(We have not explicitly shown the 20\% renormalization of the axial vector
amplitude due to QCD effects, since this percentage factor does not
significantly affect what follows.) It will be convenient to use as a measure of
the anomalous contribution the dimensionless ratio
\be
\tilde{\Delta}(E)\equiv\frac{\Delta(E)}{G_F/2\sqrt{2}}
\sim {\cal O}\left(\frac{\Delta (E)}{{\rm Re}\ C_V(0)}\right)\,.
\labeq{rdefn}
\ee
Inputting Eq.\ \req{delta} and the numerical value of $G_F$ into
\req{rdefn} gives the
useful result
\be
\tilde{\Delta}(E)\simeq\left(\frac{E/100\ \gev}{\es/10^{18}\ \mbox{eV}}\right)\
\left(\frac{\sigs}{100\ \rm{mb}}\right)\ \ .
\labeq{r}
\ee
It is clear from \eq{r}, and somewhat remarkable, that order 100\% effects in
the real elastic amplitudes begin to appear already at energies seven orders of
magnitude below the full realization of the strong cross section. This is our
main physics result, {}from which we obtain the observable consequences
discussed next.

\subsection{Anomalous Neutrino Index of Refraction and Effective Potential}
The consequence of our neutrino dispersion relation and the sensible
assumptions made for
the new high--energy neutrino interaction is the anomalous real elastic
amplitude at
lower energies given in Eq.\ \req{delta}.  The fractional increase in the
real amplitude,
compared to the SM value, is given in \req{r}.
Since the real part of the forward amplitude makes a direct contribution
to the index of refraction $n_{\rm ref}$,
the most direct test of an anomalously large neutrino cross section would be a
measurement of this refractive index.
The real part of a forward amplitude is related to the refractive index
$\index$ by
\be
n_{{\rm ref},\pm}-1=\frac{2\rho}{E}{\rm Re}A_{\pm}(E)\,,
\labeq{indx}
\ee
where $\rho$ is the nucleon number density of the (possibly polarized) medium.
The anomalous contribution to the right--hand side of \eq{indx} exceeds the SM
contribution at neutrino energies $E\,\gsim\,100$ GeV. Perhaps fortuitously,
100 GeV is roughly the mean energy of atmospheric neutrinos producing
throughgoing muons in underground detectors. According to SM physics, neutrinos
at $E=100$~GeV with mass--squared splittings $\lsim 10^{-2}$ eV$^2$ receive
significant phase contributions from matter effects, and so would receive even
larger effects from the new interaction. With $E\sim 100$~GeV and $\delta m^2
\sim 10^{-2} {\rm eV}^2$, there could be sizeable new matter--effects on
oscillations over a distance of the order of the earth diameter. However, if
the anomalous reactions (if they exist) are flavor neutral, they produce a
common phase and there will be no new matter effects associated with them.

\subsection{Low--Energy Elastic Scattering}
There are more promising observable consequences available from the elastic
cross section, obtained from the square of the elastic amplitude.
The normalization of the amplitudes is such that the spin-averaged elastic
scattering
cross section in the forward direction is given by
\be
\left.\frac{d\sigma}{dt}\right|_{t=0}\ =\ \frac{1}{\pi}\
(|C_V(E)|^2+|C_A(E)|^2)\ \
.
\labeq{dsdt}
\ee
In order to simplify the following discussion, let us
assume that the elastic form factors of the nucleon effectively cut off
at some common effective $|t|\sim\msqeff$,
and that the elastic amplitude is approximately
real for $|t|\,\lsim\,\msqeff.$
Then we may approximate the
spin-averaged neutrino-nucleon elastic cross section as
\be
\sigma^{\nu N}_{el}(E)\simeq \frac{\msqeff}{\pi}\ \left[({\rm Re}\ C_V(E))^2+
({\rm Re}\ C_A(E))^2\right]\ \ .
\labeq{sigel}
\ee

It is useful now to frame the discussion in terms of the cross--section
normalized to
the SM value:
\be
\tsig_N\equiv \frac{\sigma^{\nu N}_{el}(E)}
{\left.\sigma^{\nu N}_{el}(E)\right|_{Z-exchange}}
=\frac{\left[\tdel + a_N \right]^2\ +\ 1}{a_N^2\ +\ 1}\ \ ,
\labeq{tsig}
\ee
with
\be
a_N=4\sin^2{\theta_W}Q(N)-sgn(T_3(N))
=\left\{
\begin{array}{ll}
           4\sin^2{\theta_W}-1 & \mbox{for $N=p$}\\
           1                   & \mbox{for $N=n$}
\end{array}
\right. \,.
\labeq{rn}
\ee
Use has been made of Eqs.~\req{vadelta}, \req{vazero}, \req{rdefn} and
\req{sigel} in arriving at Eqs.~\req{tsig} and \req{rn}.
With $\sin^2{\theta_W}=0.23,$
one finds
\bea
\tsig_{\rm proton}&\simeq&\frac{(\tdel-0.080)^2+1}{(0.080)^2+1}\,,\nns
\tsig_{\rm neutron}&\simeq& \frac{(\tdel+1)^2+1}{2}\ \ .
\labeq{rprn}
\eea

{}From \eq{rprn}, it is apparent that a significant enhancement in the elastic
cross sections (say a factor of 10 or more) is obtained for $\tdel> 3.$ It is
encouraging that such a value of $\tdel$ may be within reach of current
experimental setups, rather than orders of magnitude beyond. We can see from
Eq.\ \req{r} that the anomalous elastic scattering cross--section is already
significant at energy $E$ related to the energy \es characterizing the
anomalous inelastic scattering by
\be
E \simeq \left(\frac{\tdel (E)}{3}\right)
\left(\frac{100\ \rm{mb}}{\sigs}\right)
\left(\frac{\es}{10^{17.5}{\rm eV}}\right) \times 100\,\gev\,.
\ee
This result says that if the neutrino is strongly interacting at $\es \sim
10^{17.5}{\rm eV}$, then an anomalous rise in the elastic cross--section is
occurring at neutrino energies already available at existing accelerators. The
Fermilab neutrino beam used by the NuTeV experiment has a mean energy of about
100 GeV. The CERN neutrino beam used in oscillation experiments
contains neutrinos at 100 GeV, with an intensity down by a factor of $\sim 20$
{}from the intensity at the mean energy of 30 GeV.

We propose that a comparison of the elastic event rate below and above 100 GeV
be done, to look for the onset of anomalous neutrino elastic scattering.
The proposed experiment is difficult, for the
identification of elastic neutrino scattering is challenging.
A low--energy recoil proton must be detected, with a veto on events with
pions produced.
Because the momentum transfer in elastic scattering is limited to
$\lsim 1\ \gev^2,$ the recoil nucleon has a kinetic energy of at most 0.5 GeV.

We end this section by noting that since the ratios $\tsig_{p,n}$ grow
quadratically with $E,$ anomalies in the elastic cross section develop rapidly
for $E > 100\ \gev.$
Thus, the event sample of a future underground/water/ice neutrino telescope
optimized for TeV neutrinos could conceivably contain {\em 1000} times more
elastic neutrino events than predicted by the SM;
and a telescope optimized for PeV neutrinos may contain $10^9$ more elastic
events.

\section{Discussion and Conclusions}

The hypothesized new \si\ for neutrinos with energy above \egzk\
is extraordinarily speculative.
The only phenomenological argument for the hypothesis is that it provides a
possible
explanation of the super-GZK events.
Yet it may be testable at lower energies
because it implies, as we have shown,
an anomalous contribution to the real amplitude at lower energies.
The dispersion--relation formalism presented here provides a
model-independent theoretical framework
for examining the lower energy implications of the
high--energy neutrino strong--interaction hypothesis.


We have discussed two classes of low--energy tests of the neutrino
strong--interaction hypothesis. The first uses the real amplitude directly to
calculate an anomalous neutrino index of refraction, resulting in possible
anomalous matter effects for flavor--oscillations. This approach is viable only
if there is flavor dependence in the anomalous interaction. The other class
uses the squared amplitude to calculate anomalous neutrino--nucleon elastic
scattering. With elastic scattering, the hypothesis may be testable already
using 100 GeV neutrinos, even though the strong--interaction inelastic
cross--section does not develop until near \egzk
.

We have seen that a measurable anomalous elastic signal at 100 GeV requires a
value of $10^{17.5}$ eV for $\es$, characterizing the high--energy anomaly. Is
a value of $\es$ as low as this possible? For now, the data do not rule out
such a possibility. The break in the very high energy cosmic ray data (the
``ankle'') seems to occur at an energy of $10^{18.5}$ eV, where it is thought
that the transition between galactic and extra-galactic sources is evolving.
This is not inconsistent with neutrinos attaining their hadronic cross sections
an order of magnitude lower in energy, but not yet dominating the cosmic ray
spectrum. This picture gains some support from a two-component fit to the Fly's
Eye data \cite{eye}, which suggests that the extragalactic high energy spectrum
begins to appear already at about $10^{17.5}$ eV.

There may be other tests of the strong--interaction hypothesis,
beyond those formulated here.
For example, if the neutrino develops a \si\ at high energy,
do not the electron and the other charged--lepton $SU(2)$--doublet partners
of the neutrinos also develop a similar \si?
If so, is there new physics to be sought in
\cl--nucleon scattering in the highest--energy cosmic ray air--showers?
Is there new physics in the elastic $e^\pm p$ scattering channel
at HERA energies?\footnote
{However, in the elastic $e^\pm p$ channel at HERA,
a simple calculation shows that the one-photon exchange dominates the usual
neutral current interactions at small momentum transfer to such an extent that
even at the effective lab energy of $\sim 10^5$~GeV applicable to HERA the
anomalous scattering contribution is at the 1\% level.}
If elastic scattering is enhanced at lower energies, is it not likely that
quasi--elastic scattering is also enhanced? If so, the anomalous event rate
could differ significantly from what we have calculated from just the elastic
channel.\footnote {A possible enhancement in the quasi--elastic channel cannot
be deduced from dispersion relations. A separate calculation could be made, in
principle, if the details of the new high--energy strong--interaction are
specified.}

Our proposal to look for the onset of an anomalous enhancement in the elastic
scattering rate around 100 GeV, using presently available neutrino beams, is a
conservative first step toward empirically answering these questions.

\vspace{0.5cm}\noindent
{\bf Acknowledgements:} This work was supported in part by the National Science
Foundation grant PHY-9722044 (HG), and  by the Department of Energy grant
DE--FG05--85ER40226 and the Vanderbilt University Research Council (TJW).
Some of this work was performed at the Aspen Center for Physics.
\vfill\eject

\appendix
\section{Appendix: Derivation of Neutrino Dispersion Relations}
Consider the elastic scattering of a (left-handed) neutrino, incident along the
$+z$-axis,  from a nucleon $N$ whose
mass is $M.$
The $S$-matrix can be written as
\be
S_{fi}=1+(2\pi)^4\:i\:\delta^4(P_f-P_i)\ T(k\pri,p\pri,\lm\pri;k,p,\lm)\ \ ,
\labeq{s}
\ee
where
\bea
T(k\pri,p\pri,\lm\pri;k,p,\lm)&=
&\frac{1}{(2E\:2E\pri\:2\omega\:2\omega\pri)^{1/2}}\
{\cal M}(k\pri,p\pri,\lm\pri;k,p,\lm)\ \
,\nns
{\cal M}(k\pri,p\pri,\lm\pri;k,p,\lm)&=
&\bar u_{\nu}(k\pri)\gamma^{\mu}(1-\gamma_5)u_{\nu}(k)\nns
&&\ \cdot\
\bar u_{N}(p\pri,\lm\pri)\gamma_{\mu}\left[A_-(s,t)L+A_+(s,t)R\
\right]u_{N}(p,\lm)\ \ ,
\labeq{t}
\eea
with $\lm,\lm\pri$ labeling the initial and final nucleon helicities, and the
projection operators $L,R=(1\mp\gamma_5)/2$;
volume factors in the normalization of $T$ are omitted.
The Mandelstam variables are defined as usual:
\bea
s&=&(k+p)^2=(k\pri+p\pri)^2\nn\\
t&=&(k-k\pri)^2=(p-p\pri)^2\nn\\
u&=&(p-k\pri)^2=(p\pri-k)^2\nn\\
\labeq{mand}
\eea
with $s+t+u=2M^2+2\mnus.$ (We retain a small neutrino mass $\mnu$ for the
moment.)

The optical theorem relating the forward amplitude to the total cross
section (for a fixed initial nucleon spin) reads
\be
{\rm Im}\ {\cal M}(k,p,\lm ;k,p,\lm)= 2 M\sqrt{E^2-\mnus}
\ \sigma^{\nu N}_{tot}(E,\lm)\ \ ,
\labeq{opt}
\ee
where $E$ is the neutrino lab energy and $\lm$ is now the nucleon spin
along the $z$-axis.
A little bit of algebra shows that for forward scattering

\be
{\cal M}(k,p,\pm\ ;k,p,\pm)=8ME\ \  \apm(s,0)\ \ .
\labeq{ta}
\ee

It will prove convenient to
work in terms of the invariant quantity $\nu,$
defined by

\bea
\nu&\equiv& (p+p\pri)\cdot(k+k\pri)/(4M)\nns
&=&(s-u)/(4M)\ \ .
\labeq{nu}
\eea
For forward elastic scattering of a neutrino of lab energy $E$ on a stationary
target of mass $M,$
$\nu=E.$

Ignoring subtractions for the
moment, the analytic property of $\apm$ is expressed through the Hilbert
transform
in the $\nu$-plane (for fixed $t)$
\be
\apm(\nu,t)=\frac{1}{\pi}\int_{-\infty}^{+\infty}\ \frac{d\np\ {\rm
Im}\apm(\np,t)}
{\np-\nu-i\epsilon}\ \ .
\labeq{hilbert}
\ee
 There are two cuts:
\bi
\itm{1}$s$-channel cut -- from $s=(M+\mnu)^2$ to $s=\infty$ from physical
$\nu F$
scattering. In the $\nu$-plane
this gives a cut from $\nuth=\mnu+t/(4M)$
to $\nu=\infty.$
\itm{2}$u$-channel cut -- from $u=(M+\mnu)^2$ to $u=\infty$ from physical
$\bar\nu N$ scattering. Substituting
$s=2M^2+2\mnus-u-t$ into \eq{nu} we find that in the $\nu$-plane the
$u$-channel cut extends from $\nu=-\infty$ to $\nu=-\nuth$.\ei
Thus,
\eq{hilbert} becomes

\be
\apm(\nu,t)=\frac{1}{\pi}\int_{-\infty}^{-\nuth}\ \frac{d\np\ {\rm Im}\
\apm(\np,t)}
{\np-\nu-i\epsilon}\;\; + \;\; \frac{1}{\pi}\int_{\nuth}^\infty\ \frac{d\np\
{\rm Im}\ \apm(\np,t)}
{\np-\nu-i\epsilon}\ \ .
\labeq{disp}
\ee
\smallskip

\noindent
With a change in variables $\np\rightarrow -\np$ in the second integral,
\eq{disp} becomes
\be
\apm(\nu,t)=\frac{1}{\pi}\int_{\nuth}^{\infty}\ \frac{d\np\ {\rm Im}\
\apm(\np,t)}
{\np-\nu-i\epsilon}\;\; -\;\; \frac{1}{\pi}\int_{\nuth}^\infty\ \frac{d\np\
{\rm Im}\ \apm(-\np,t)}
{\np+\nu+i\epsilon}\ \ .
\labeq{dispa}
\ee
We now use crossing. First define an amplitude for $\bar\nu N$ scattering
in a manner
analogous to \eq{t}:

\bea
\ov{{\cal M}}(k\pri,p\pri,\lm\pri;k,p,\lm)&=&\bar
v_{\nu}(k)\gamma^{\mu}(1-\gamma_5)v_{\nu}(k\pri)\nns
&&\cdot\
\bar u_{N}(p\pri,\lm\pri)\gamma_{\mu}\left[\bar A_-(\nu,t)L+\bar A_+(\nu,t)R
\right]u_{N}(p,\lm)\ \
\labeq{tbar}
\eea

It is a lengthy but straightforward exercise to derive the crossing relation
between  $\apm$ and the analogous amplitude for the right--handed antineutrino,
$\bar \apm$. The method used is essentially that which can
be found in \cite{goldberger}, and can also be checked by constructing an
effective Lagrangian which will yield the amplitude \req{t} at tree level. The
result is
\be
{\rm Re}\ \bar \apm(\nu,t)= - {\rm Re}\ \apm(-\nu,t)
\labeq{xre}
\ee
The minus sign comes from the anticommuting properties of fermion
operators, which come into play when the antineutrino in- and out-field
operators are
brought into normal order.
By writing a dispersion relation for  $\bar\apm$ similar to \eq{dispa}, and
comparing to
the latter using \eq{xre}, one discovers that consistency requires
\be
{\rm  Im}\ \bar  \apm(\nu,t)= +\ {\rm Im}\ \apm(-\nu,t)
\labeq{xim}
\ee
Substituting \req{xim} into \req{dispa}, one obtains

\be
{\rm Re}\ \apm(\nu,t)=\frac{1}{\pi}{\cal P}\int_{\nuth}^{\infty}\ d\np\ \left(
\frac{{\rm  Im}\ \apm(\np,t)}{\np-\nu}\ -\ \frac{{\rm  Im}\ \bar
\apm(\np,t)}{\np+\nu}
\right)\ \ ,
\labeq{dispb}
\ee
with ${\cal P}$ signifying the principal value of the integral.

Now specialize to the forward direction $(t=0)$ and to massless neutrinos
(in which case $\nuth=0).$ We will call the forward (invariant) scattering
amplitude $\apm(E),$ reverting to the neutrino lab energy as the kinematic
variable.
With the use
of Eqs.~\req{opt} and \req{ta}, the optical theorem in terms of the
amplitude $\apm$
reads
\be
{\rm Im}\  \apm(E)=\tfrac{1}{4}
\ \sigma^{\nu N}_{tot}(E,\pm)\ \ ,
\labeq{opta}
\ee
with a similar equation for $\bar \apm:$
\be
{\rm Im}\ \bar \apm(E)=\tfrac{1}{4}
\ \sigma^{\bar \nu N}_{tot}(E,\pm)\ \ .
\labeq{optabar}
\ee
After inserting Eqs.\ \req{opta} and \req{optabar},
the dispersion relation \req{dispb} reads
\be
{\rm Re}\ \apm(E)=\frac{1}{4\pi}{\cal P}\int_0^{\infty}\ dE\pri\
\left(
\frac{\sigma^{\nu N}_{tot}(E\pri,\pm)}{E\pri-E}\ -\
\frac{\sigma^{\bar \nu N}_{tot}(E\pri\pm)}{E\pri+E}
\right)\ \ .
\labeq{dispc}
\ee
To improve the convergence of the integral,\footnote{Note that the Pomeranchuk
theorem does {\em not} hold in the electroweak theory -- if it did, the
threshold amplitude ${\rm Re}\ \apm(0)$ could be calculated from \req{dispc}.}
we rewrite
\req{dispc} in the once-subtracted form:

\be
{\rm Re}\ \apm(E)-{\rm Re}\ \apm(0)=\frac{E}{4\pi}{\cal P}\int_0^{\infty}\
dE\pri\
\left(
\frac{\sigma^{\nu N}_{tot}(E\pri,\pm)}{E\pri(E\pri-E)}\ +\
\frac{\sigma^{\bar \nu N}_{tot}(E\pri,\pm)}{E\pri(E\pri+E)}
\right)\ \ .
\labeq{App:dispsubt}
\ee
No information is lost in the subtraction, since the subtraction constant
${\rm Re}\ \apm(0)$ is known from the standard model.
\eq{App:dispsubt} provides the theoretical basis for the phenomenological
considerations in the main text.

\vfill\eject
\begin{thebibliography}{99}

\bibitem{akeno}
    M. Takeda et al., {\sl Phys. Rev. Lett.} {\bf 81}, 1163 (1998);
    S.Yoshida, et al., {\sl Astropart., Phys.} {\bf 3}, 105 (1995);
    N. Hayashida et al., {\sl Phys. Rev. Lett.} {\bf 73}, 3491 (1994).

\bibitem{eye} D. J. Bird et al., (Fly's Eye Collab.)
   {\sl  Phys. Rev. Lett.} {\bf 71}, 3401 (1993);
   {\sl Astrophys. J.} {\bf 424}, 491 (1994);
   {\sl ibid.} {\bf 441}, 144 (1995).

\bibitem{hp} G. Brooke et al. (Haverah Park Collab.),
   Proc. 19th Intl. Cosmic Ray Conf. (La Jolla) {\bf 2}, 150 (1985);
   reported in M. A. Lawrence, R. J. O. Reid, and A. A. Watson
    (Haverah Park Collab.),
   {\sl J. Phys. G} {\bf 17}, 733 (1991).

\bibitem{yak} N. N. Efimov et al., (Yakutsk Collab.)
   ICRR Symposium on Astrophysical Aspects
   of the Most Energetic Cosmic  Rays, ed. N. Nagano and F. Takahara,
   World Scientific pub. (1991);
   and Proc. 22nd ICRC, Dublin (1991).

\bibitem{gzk} K. Greisen, {\sl Phys. Rev. Lett.} {\bf 16}, 748 (1966);
   G. T. Zatsepin and V. A. Kuzmin,
   {\sl Pisma Zh. Eksp. Teor. Fiz.} {\bf 4}, 114 (1966);
   F. W. Stecker, {\sl Phys. Rev. Lett.} {\bf 21}, 1016 (1968);
    J. L. Puget, F. W. Stecker and J. H. Bredekamp,
        {\sl Astrophys. J.}, {\bf 205}, 638 (1976);
    V. S. Berezinsky and S. I. Grigoreva,
        {\sl Astron. \& Astrophys.}, {\bf199}, 1 (1988).

\bibitem{bzinsky}
See, e.g., {\sl Astrophysics of Cosmic Rays},
   V.S. Berezinskii et al., North--Holland pub., 1990;
   {\sl High Energy Astrophysics}, M.S. Longair, Cambridge U. pub., 2nd
ed., 1994.

\bibitem{stanev}  T. Stanev, P. Biermann, J. Lloyd-Evans, J. Rachen and
  A. Watson, {\sl Phys. Rev. Lett.} {\bf 75}, 3056 (1995).

\bibitem{AGNmax}
         L. Nellen, K. Mannheim and P. Biermann, {\sl Phys. Rev.} {\bf
D47}, 5270 (1993);
  F. W. Stecker and M. H. Salamon, {\sl Space Sci. Rev.} {\bf 75}, 341 (1996);
  K. Mannheim, {\sl Astropart. Phys.} {\bf 3}, 295 (1995);
                R. J. Protheroe, astro-ph/9607165, and astro-ph/9612213;
  W. Bednarek and R. J. Protheroe, astro-ph/9802288;
  G. C. Hill, {\sl Astropart. Phys.} {\bf 6}, 215 (1997);
  F. Halzen and W. E. Zas, {\sl Astrophys. J.} {\bf 488}, 669 (1997);
an excellent overview of models proposed to generate ultrahigh
   energy primaries is given by
        P. L. Biermann, {\sl J. Phys. G:} {\bf 23}, 1 (1997).

\bibitem{GRBmax}
   B. Paczynski and G. Xu, {\sl Astrophys. J.} {\bf 427}, 708 (1994);
          R. Plaga, {\sl Astrophys. J.} {\bf 424}, L9 (1994);
   T.J. Weiler, J.G. Learned, S. Pakvasa, and W. Simmons, hep--ph/9411432
(1994);
   F. Halzen and G. Jaczko, {\sl Phys. Rev.} {\bf D54}, 2779 (1996);
   S. Lee, {\sl Phys. Rev.} {\bf D58}, 043004 (1998);
   J. Bahcall and E. Waxman, {\sl Phys. Rev. Lett.} {\bf 78}, 2292 (1997);
      and hep--ph/9807282, to appear in {\sl Phys. Rev.} {\bf D};
   J.P. Rachen and P. Meszaros, astro--ph/9802280;
   M. Vietri, {\sl Phys. Rev. Lett.} {\bf 80}, 3690 (1998);
      and astro--ph/9806110, to appear in {\sl Astrophys. J.}

\bibitem{50Mpc}
    S. Yoshida and M. Teshima, {\sl Prog. Theor. Phys.}{\bf 89}, 833 (1993);
    F. A. Aharonian and J. W. Cronin, {\sl Phys. Rev.} {\bf D50}, 1892 (1994);
   J. W. Elbert and P. Sommers, {\sl Astrophys. J.} {\bf 441}, 151 (1995);
     S. Lee, in ref. \cite{GRBmax}.

\bibitem{nucleus} F. W. Stecker, {\sl Phys. Rev.} {\bf 180}, 1264 (1969);
   L.N. Epele and E. Roulet, astro--ph/9808104;
   F.W. Stecker and M.H. Salamon, astro--ph/9808110, {\sl Astrophys. J.},
to appear.

\bibitem{hs} F. Halzen, R. A. Vazquez, T. Stanev, and V. P. Vankov,
  {\sl Astropart. Phys.}, {\bf 3}, 151 (1995).

\bibitem{AGASApairs} N. Hayashida et al. (AGASA collaboration),
   Phys. Rev. Lett {\bf 77}, 1000 (1996).

\bibitem{BF98} P. Biermann and G. Farrar, astro-ph/9806242,
   to appear in Phys. Rev. Lett. (1998).

\bibitem{SInu} G. Domokos and S. Nussinov,
        {\sl Phys. Lett.} {\bf B187}, 372 (1987);
   G. Domokos and S. Kovesi--Domokos, {\sl Phys. Rev.} {\bf D38}, 2833 (1988);
   J. Bordes,  Hong--Mo Chan, J. Faridani, J. Pfaudler, and T. S. Tsun,
        hep--ph/9705463; Astropart. Phys. {\bf 8}, 135 (1998).

\bibitem{bgh98}
   G. Burdman, F. Halzen, and R. Gandhi, Phys. Lett. {\bf B417}, 107 (1998).

\bibitem{meanmo}
J. Bordes,  Hong--Mo Chan, J. Faridani, J. Pfaudler, and T. S. Tsun,
hep--ph/9711438.
%
%
\bibitem{goldberger}
M.L. Goldberger , M.T. Grisaru , S.W. MacDowell, and D.Y. Wong
{\em Phys.~Rev.}~{\bf 120} (1960) 2250.
\eb
\ed